\documentclass[12pt]{article}
\usepackage{latexsym}
\usepackage{amsmath}
\usepackage{indentfirst}
\usepackage{cite} 

\long\def\ca#1\cb{} 

\newcommand{\becs}{\begin{cases}}
\newcommand{\bem}{\begin{matrix}}

\newcommand{\encs}{\end{cases}}
\newcommand{\enm}{\end{matrix}}

\newcommand{\ket}[1]{|#1\rangle }

\newcommand{\lra}{\leftrightarrow }

\newcommand{\EC}{{\mathcal E}}
\newcommand{\FC}{{\mathcal F}}

\newcommand{\lm}{\lambda }

\newcommand{\om}{\omega }

\def\outl#1{\par{\medskip\noindent\hspace*{0.1cm}\bf
      \mathversion{bold}#1\mathversion{normal}\smallskip} }
   \def\xa{} \def\xb{}  

 \def\outl#1{}\def\xa{}\def\xb{}

\ca
 \def\outl#1{\par{\medskip\noindent\hspace*{.5cm}\bf
      \mathversion{bold}#1\mathversion{normal}\smallskip} }
 \long\def\xa#1\xb{} 
  
\cb

\begin{document}

\title{Reply to ``Comment on `Nonlocality claims are inconsistent with
  Hilbert-space quantum mechanics' ''}
\author{Robert B. Griffiths\thanks{Electronic address: rgrif@cmu.edu}\\
  Department of Physics\\
  Carnegie Mellon University\\
  Pittsburgh, PA 15213}

\date{Version of 21 Dec. 2021}
\maketitle

\xa
\begin{abstract}
  In Phys. Rev. A 101, 022117 (2020), it was argued that Bell inequalities are
  based on classical, not quantum, physics, and hence their violation in
  experiments provides no support for the claimed existence of peculiar
  nonlocal and superluminal influences in the real (quantum) world. This Reply
  to Lambare's Comment, Phys. Rev. A 104, 066201 (2021), on that paper seeks to
  clarify some issues related to the correct use of Hilbert space quantum
  mechanics for identifying the microscopic causes of later macroscopic
  measurement outcomes, a matter not properly addressed by Bell, who used
  classical hidden variables in place of the Hilbert subspaces (equivalently,
  their projectors) employed by von Neumann in his \it{Mathematical Foundations
    of Quantum Mechanics}.
\end{abstract}

\xb
\tableofcontents
\xa

\xb
\section{Introduction \label{sct1}}
\xa

\xb
\outl{HSQM. Property$\lra$=Projector.  HSQM includes stochastic time
  development }
\xa

Placing Ref.~\cite{Grff20} in a historical context will assist in
responding to Lambare's Comment\cite{Lmbr21}. As its title suggests, the
argument in Nonloc is based upon \emph{Hilbert space quantum mechanics} (HSQM),
that is to say the basic framework laid down by von Neumann (who invented the
term `Hilbert space') in \cite{vNmn32b}: a complex vector space with an inner
product---we may assume it to be finite-dimensional for the present
discussion---with physical properties represented by its subspaces or by the
corresponding projectors (orthogonal projection operators) onto these
subspaces. For example, in the case of a harmonic oscillator the quantum
\emph{property} that the energy is not greater than $2\hbar\om$ is represented
by the subspace spanned by its two lowest eigenstates, or by the rank 2
projector onto this subspace. Since von Neumann considered Born's probabilistic
interpretation an essential component of quantum theory, stochastic
(probabilistic) time development can also be considered a part of HSQM.

\xb
\outl{CH based on but extends HSQM}
\xa

The \emph{consistent histories} (CH) interpretation of quantum
theory\footnote{%
  See \cite{Grff19b} for an overview. Both \cite{Grff17b} and \cite{Grff20}
  provide short summaries of essential ideas, which are spelled out in much
  greater detail in \cite{Grff02c}.}
used in \cite{Grff20} and this Reply, is based firmly upon HSQM, but includes
additional ideas, notably that of a \emph{quantum history}: a sequence of
quantum properties at a succession of times. It allows a probabilistic
description of quantum time development analogous to a classical stochastic
process, with no need to refer to measurements. From the CH perspective
measurements are simply particular instances of quantum processes governed by
fundamental quantum principles that make no reference to measurements.

\xb
\outl{Noncommutation. Bell ignored it. EPRB. Experiments agreed with HSQM.
Bell's mistaken nonlocality claims. Outline of remainder of Reply}
\xa

The central feature that distinguishes HSQM from classical physics is that two
projectors $P$ and $Q$ representing different properties \emph{need not
  commute}: $PQ$ can be different from $QP$. Bell when deriving his famous
inequalities \cite{Bll64b} ignored quantum noncommutation and assumed that
microscopic quantum properties could be represented by \emph{classical}, i.e.,
\emph{commuting} hidden variables, something von Neumann had rejected. When
experiments related to Bohm's version, Ch.~22 of \cite{Bhm51}, of the
Einstein-Podolsky-Rosen paradox \cite{EnPr35}---we abbreviate this as
EPRB---were carried out to test the Clauser, Horne, Shimony, and Holt (CHSH)
version\cite{CHSH69} of a Bell inequality, the results agreed with HSQM and
violated the inequality. Bell, followed by many others, ascribed this
disagreement to the presence in the quantum world of mysterious nonlocal
influences. While various objections were raised at that time and later, it was
the introduction of the CH approach in the 1980s and its later development,
continuing for some years after Bell's untimely death, that has allowed a
detailed understanding of the flaws in Bell's approach, and how to replace it
with a correct quantum analysis that eliminated the need for ``spooky
nonlocality''. The most recent and detailed presentation is in \cite{Grff20}.

\xb
\outl{Reply to Comment use same notation for EPRB. Outline of Reply}
\xa

It is hoped that this Reply to Lambare's Comment will assist other readers in
better understanding some of the material in \cite{Grff20}, in particular how to
identify microscopic \emph{causes} of later macroscopic measurement outcomes.
Both Comment and Reply employ the same spin-half notation for EPRB. In
particular, $A$ and $B$, which take values $\pm1$, represent the macroscopic
outcomes of measurements by Alice and Bob using measurement settings $a$ and
$b$; e.g., $a=x$ means that Alice measures $S_x$. Section~\ref{sct2} is a very
brief introduction to some simple ideas about \emph{causes}; Sec.~\ref{sct3} is
a short discussion of how to use them for discussing spin measurements; and
Sec.~\ref{sct4} is a reply to various issues raised in the Comment. It is
followed by a brief conclusion in Sec.~\ref{sct5}.

\section{Causes \label{sct2}}

\xb
\outl{Intuitive idea of a 'cause'}
\xa

The intuitive idea of a \emph{causal relationship} between events $F$ and $G$
at two times $t_1<t_2$ in a probabilistic theory can be thought of in the
following way. They are \emph{statistically independent} provided the joint
probability distribution factors,
\begin{equation}
 \Pr(F,G) = \Pr(F)\Pr(G),
\label{eqn1}
\end{equation}
and otherwise they are \emph{correlated}, which means that
\begin{equation}
 \Pr(G|F) = \Pr(F,G)/Pr(F) \neq \Pr(G),
\label{eqn2}
\end{equation}
so the probability of the later event $G$ depends in some way upon whether $F$
did or did not occur earlier, thus suggesting that $F$ may somehow have
influenced $G$. The strongest possible correlation, which would characterize a
\emph{complete} or \emph{ideal} cause, is:
\begin{equation}
 \Pr(G|F) = 1,\quad \Pr(F|G)=1.
\label{eqn3}
\end{equation}
That is, if $F$ occurs one can be sure $G$ will occur later, and if $G$ occurs,
it was surely preceded by $F$. The complete \emph{absence} of a cause is
when $F$ and $G$ are statistically independent, \eqref{eqn1}.

\xb
\outl{Common cause}
\xa

But there are situations in which \eqref{eqn3} holds, and yet one would not say
that $F$ caused $G$. In particular, if there is an event $E$ at a time earlier
than either $F$ or $G$, and it is an ideal cause of both $F$ and $G$ in the
sense that $\Pr(F|E) = 1 = \Pr(E|F)$ and $\Pr(G|E) = 1 = \Pr(E|G)$, $E$
could be the \emph{common cause} of both $F$ and $G$, neither of which is the
cause of the other. For example, let $E$ be a signal Charlie sends to both
Alice and Bob, who receive it as $F$ and $G$, respectively. Obviously what
Alice receives is not the cause of what Bob receives, even if it arrives
earlier, and vice versa.

\xb
\outl{Preceding only sketches the general idea of a cause. PROBABILISTIC MODEL
needed for discussing cause. In CH this is a FRAMEWORK. Single framework rule:
Incompatible frameworks cannot be combined.}
\xa

These brief remarks are intended to help orient the following discussion, and
in no sense constitute a complete theory of causes. Conditional probabilities
might be be less than 1; $E$, $F$, and $G$ could be quantities taking on a
number of different values, etc. Of particular importance is the fact
that \emph{causes cannot be discussed in this manner without a well-defined
  probabilistic model}. In the CH approach a probabilistic sample space or
\emph{framework} is a collection of commuting projectors that sum to the
identity. If some projectors in one framework do not commute with projectors in
a second framework the two frameworks are \emph{incompatible}, and
probabilistic reasoning based on one cannot be combined with that based on the
other, an instance of the \emph{single framework rule}. Further details will be
found in \cite{Grff11b,Grff14,Grff19b}.

\section{Measurements \label{sct3}}

\xb
\outl{Laboratory measurement of particle decay}
\xa

Consider a laboratory setup in which a detector detects a particle, perhaps an
alpha particle or gamma ray, coming from the decay of a radioactive source. The
experimenter might interpret the detector click as \emph{caused} by the
particle passing through a small hole of diameter 2 mm in a thick metal
collimator placed just in front of the detector. Observing that the detector
never triggers if the hole is blocked, and always detects particles
deliberately sent through the hole in separate calibration runs, supports this
notion of a cause. But a textbook discussion using unitary time evolution until
the detector clicks, completing a measurement, runs into the difficulty that
Schr\"odinger's equation applied to the spherical wave of the alpha particle
just after the decay cannot shrink it to a narrow wavepacket that can pass
through the collimator hole \emph{before} it reaches the detector. The same
problem is present in the case of a gamma ray.

\xb
\outl{Spin-half measurement setup with successive times}
\xa

The CH approach to such a situation is best explained using the much simpler
case of measuring the spin of a spin-half particle. Let the particle be
prepared in some spin state at time $t_0$ and then travel undisturbed through a
region with no magnetic field until a time $t_1$ just before it interacts with
Alice's spin measurement apparatus with setting $a$, resulting in a macroscopic
outcome $A = \pm 1$ at a later time $t_2$. If $a=z$, so a spin $S_z= +1/2$, in
units of $\hbar$, at $t_1$ leads to the later outcome $A=+1$, and $S_z= -1/2$
to $A=-1$, is Alice justified in identifying $S_z$ at $t_1$ as the \emph{cause}
of the later $A$ outcome? We assume she has checked her apparatus using
calibration runs in which particles with known values of $S_z$ sent into it
resulted in the corresponding $A$ outcomes.

\xb
\outl{$S_z$ measurements using $\FC_z$ framework; $\EC_{xz}$ includes earlier
  $S_x$}
\xa

Let $\FC_z$ be a probabilistic sample space, in CH terminology a
\emph{framework}, of four quantum histories with $S_z=\pm 1/2$ at $t_1$
followed by $A=\pm 1$ at $t_2$, and probabilities assigned using the Born
rule. Since $S_z$ at $t_1$ is perfectly correlated with $A$ at $t_2$, it can be
considered the microscopic \emph{cause} of the later measurement outcome. This
conclusion is not altered if the $\FC_z$ framework is refined to make the
framework $\EC_{xz}$, a collection of 8 histories obtained by adding to each of
the histories in $\FC_z$ the two possibilities $S_x=\pm 1/2$ at the time $t_0$
when the particle was prepared. The CH analysis using the $\EC_{xz}$ framework
again leads to the conclusion that the $S_z$ value at $t_1$ is perfectly
correlated with $A$ at $t_2$, and is thus the cause of the latter.

\xb
\outl{$\EC_{xx}$: $S_x$ at $t_0$ and $t_1$}
\xa

However, there is an alternative framework $\EC_{xx}$ in which the $S_z$ values
$\EC_{xz}$ at $t_1$ are replaced with $S_x$ values: so one has $S_x$ values at
both $t_0$ and $t_1$, followed by $A$ values at $t_2$. (Note that we are
continuing to examine the case in which Alice's measurement setting is $a=z$.)
The $\EC_{xx}$ framework is incompatible with the $\EC_{xz}$ framework, since
$S_x$ and $S_z$ at $t_1$ do not commute. Using $\EC_{xx}$ one can show that
the $S_x$ values $\pm 1/2$ at $t_1$ are statistically independent of, so
cannot be thought of causing, the later outcome $A$. (Note that
the $\EC_{xx}$ framework is the one students are taught in textbook quantum
mechanics: unitary time evolution up until the measurement begins, followed by
a mysterious ``collapse''. It is of little help in identifying the microscopic
causes of laboratory measurement outcomes.)

\xb
\outl{Framework is chosen by physicist analyzing data; choice does \emph{not}
  influence the physical process}
\xa

The choice of \emph{which} framework to use is made by the physicist when
analyzing experimental data in order to understand its physical significance,
and has no influence on the actual physical process. For experimental runs with
an initial $S_x$ preparation and measurement setting $a=z$, Alice can use
either $\EC_{xz}$ or $\EC_{xx}$. But it is only the former that allows her to
identify a \emph{cause} for the later $A$ outcome. And since the two frameworks
are incompatible the corresponding conclusions cannot be combined. Thus given
an initial $S_x=-1/2$ at $t_0$ and a final $A=+1$ at $t_2$, Alice can use
either $\EC_{xx}$ to infer $S_x=-1/2$ at $t_1$, or $\EC_{xz}$ to infer
$S_z=+1/2$ at $t_1$. But it makes no sense to combine them, in violation of the
single framework rule, to conclude that \emph{both} $S_x=-1/2$ \emph{and}
$S_z=+1/2$ were simultaneously true at $t_1$. There is no projector in the
spin-half Hilbert space that can represent such a combination.

\xb
\outl{Change $a=z$ to $a=x$ setting before $t_1$ does not influence particle}
\xa

If at any time prior to $t_1$ Alice changes the $a=z$ setting of her apparatus
to $a=y$ in order to measure $S_y$, she can use an $\FC_y$ framework, $S_y$
values at $t_1$ followed by $A$ values at $t_2$, in order to infer an earlier
$S_y$ value from the later outcome $A$. Setting $a=y$ rather than $a=z$ does
not somehow ``bring into existence'' an $S_y$ value at $t_1$; instead it allows
Alice to learn a particular feature about the past, something she could not
have learned using the alternative $a=z$ setting. Classical intuition might say
that because in a particular run Alice is at liberty to choose either an $a=z$
or an $a=y$ setting, therefore in this run the particle had both $S_z$ and
$S_y$ values at $t_1$. Such classical intuition applied in a quantum context
can be badly misleading. And just as Alice's choice cannot influence the
earlier state of the particle, the earlier state of the particle cannot
influence Alice's choice. Her choice could be made by flipping a coin (or a
quantum coin, Sec.~19.2 of \cite{Grff02c}). When properly understood, quantum
mechanics is a \emph{local} theory.

\section{Reply to Comment \label{sct4}}

\xb
\outl{Reply addresses main criticisms. Focus is EPRB and CHSH}
\xa

Various criticisms of \cite{Grff20} are found in different sections of the
Comment. This Reply responds to what appear to be the main objections, without
attempting to respond to every statement in detail. Note that numbered
equation $N$ in the Comment is referred to below as `Eq.~($N$)', whereas
equations in this Reply are referenced without the preceding `Eq.'

\subsection{ Different Routes to CHSH \label{sbct4.1}}

\xb
\outl{Alternative derivations of the wrong answer given Comment Sec.~III}
\xa

In his Sec.~III, Lambare points out that while a particular Bell inequality
derivation---it is primarily CHSH that is in view---might fail due to faulty
premisses or bad reasoning, this does not exclude the possibility that a
\emph{different} derivation could lead to the \emph{same} inequality. This is
true but of little consequence for quantum physics, since such alternative
derivations only lead to what most physicists believe to be the wrong answer:
inequalities that disagree with the outcomes of numerous experiments, whose
results confirm the correctness of the correlations calculated using HSQM. If a
student turns in the wrong answer on an examination it may be difficult to
locate the mistake in his reasoning, but the answer is still wrong. In Sec.~III
Lambare notes that ignoring noncommutation, as pointed out by Khrennikov, is
one possible mistake. Another is the assumption of a joint probability
distribution that does not exist in HSQM because it involves noncommuting
projectors. The argument in Eq.~(4) of the Comment, involving Bell's notion of
local causality (BLC) and the statistical independence of measurement
choices, fails because BLC is inconsistent with quantum physics, as discussed
in Sec.~\ref{sbct4.2} below.

\xb
\outl{Parameter and Outcome Independence}
\xa

In Sec.~II of the Comment, Lambare states that a combination of \emph{Parameter
  Independence} and \emph{Outcome Independence}, terms used by Jarrett and
Shimony, lead to the CHSH inequality, so it is may be useful to locate the
source of this mistake. `Parameter Independence' means that the probability of
Alice's outcome $A$ does not depend upon Bob's choice of measurement setting
$b$; likewise $B$ is statistically independent of $a$. Hence choosing the
parameter $b$ does not allow Bob to signal Alice, and choosing $a$ does not
allow Alice to signal Bob. In \cite{Grff20} this absence of signaling is a
\emph{consequence} of a correct quantum analysis, not an independent
assumption, as some of the remarks in the Comment might seem to suggest. Thus
quantum theory confirms Parameter Independence. 'Outcome Independence', on the
other hand, means that it is possible to identify a \emph{common cause} for the
correlations of the macroscopic outcomes in the EPRB situation. Since a proper
quantum analysis \emph{does} identify such a common cause, see Sec.~V of \cite{Grff20}
and Sec.~\ref{sbct4.2} below, it is the incorrect assumption that the common
cause must be \emph{classical} that leads to the incorrect CHSH inequality.

\subsection{Bell's local causality and quantum common causes\label{sbct4.2}}

\xb
\outl{Lambare agrees his Eq. (6), \emph{factorization condition}, classical} 
\xa

\xb
\outl{Lambare Eq. (7) as Bell's  Locality Condition}
\xa

In his Sec.~IV Lambare agrees with \cite{Grff20} that the \emph{factorization
  condition}, Eq.~(6) in the Comment and Eq.~(24) in \cite{Grff20}, a central
assumption in many derivations of Bell inequalities, is based on classical
rather than quantum physics, indicating at least one reason why these
inequalities are violated in the real world. But then he asserts
there is another way of arriving at a contradiction between quantum theory and
locality:  Bell's notion of local causality leads to a formula,
\begin{equation} 
\Pr(A,B|a,b,\lm) = \Pr(A|a,\lm)\Pr(B|b,\lm),
\label{eqn4}
\end{equation} 
which is Eq.~(7) in the Comment. Lambare goes on to claim that this leads to
a contradiction, Eq.~(9), when in his Eq.~(8)  $\lm$ is equated to the
quantum singlet state $\ket{\psi}$ of two spin-half particles, for outcomes
$A=B=1$ and measurement settings $b=a$.

\xb
\outl{\emph{Intuitive} notion of local causality is correct; 
see Sec.~\ref{sct3}}
\xa

In response it is worth noting that Bell's starting point was a very plausible
\emph{intuitive} notion of \emph{local causality}: If a cause can be found for
an event $G$ occurring at a particular location at a particular time, it is
plausible that one should be able to identify a cause $F$ in its recent past
and nearby in space. When $G$ is Alice's outcome $A$ in the EPRB scenario, the
argument in \cite{Grff20} shows that this intuition is correct: the local cause
is the microscopic property of the particle just before it reaches her
apparatus. See in addition the discussion above in Sec.~\ref{sct3}.

\xb
\outl{\eqref{eqn4} satisfied if $\lm$ depends appropriately on $a$ and $b$}
\xa

\xb
\outl{Additional $\Pr(a,b) = \Pr(a)\Pr(b)$ $\Pr(\lm|a,b) = \Pr(\lm)$ used by
  Lambare}
\xa

Next note that \eqref{eqn4}, Eq.~(7) in the Comment, is satisfied in the EPRB
situation discussed in Sec.~V~C of \cite{Grff20} when $\lm$ has an appropriate
dependence upon $a$ and $b$, something not excluded in \eqref{eqn4} the way it
is written. In discussions based on classical hidden variables it is customary
to supplement \eqref{eqn4} with two additional conditions:
\begin{align}
  &\Pr(a,b) = \Pr(a)\Pr(b),
\label{eqn5}\\
  &\Pr(\lm|a,b) = \Pr(\lm),
\label{eqn6}
\end{align}
where \eqref{eqn6} is the same as Eq.~(10) in the Comment. Both \eqref{eqn5}
and \eqref{eqn6} are assumed in the Comment, though not stated explicitly, in
the course of moving from Eq.~(7) to Eq.~(8) before arriving at Eq.~(9), which
demonstrates that this line of reasoning leads to an incorrect conclusion.

\xb
\outl{Lambare error: \emph{single} earlier cause  $\lm$ for all $a$, $b$}
\xa

The fundamental difficulty with Lambare's analysis is the assumption that one
can identify a \emph{single} quantum cause $\lm$ for the outcomes of
measurements using \emph{different} measurement settings $a$ and $b$. That this
reasoning is incompatible with quantum physics when different settings
correspond to measuring incompatible spin components was discussed above in
Sec.~\ref{sct3}. Thus Bell's mistake was not in his intuitive idea of a local
cause, confirmed by a consistent quantum analysis, but its mathematical
embodiment using classical reasoning incompatible with HSQM. Equating this
\emph{single} $\lm$ to an initial quantum state $\ket{\psi}$ may help conceal
but does nothing to remedy this fundamental difficulty.

\xb
\outl{Preceding remarks also address Sec,~V, Qm Common Cause, of Comment}
\xa

The above remarks should in addition suffice to address the concerns about
\emph{quantum common causes} in Sec.~V of the Comment, as they indicate the
flaw in its Eq.~(10), identical to \eqref{eqn6} above. A discussion of
\emph{quantum} causes, including \emph{quantum common causes}, must be based
upon HSQM, \emph{not} classical physics.

\section{Conclusion \label{sct5}}

\xb
\outl{Cannot discuss Qm causes using Cl physics}
\xa

\xb
\outl{Bell inequalities useful in showing that ideas disagreeing with HSQM
can lead to results in disagreement with experiments.}
\xa

\xb
\outl{Proper way to honor Bell: apply his critical attitude towards arm waving
  and sloppy thinking, as in his `Against Measurement'}
\xa

It is hoped this Reply has adequately addressed the issues raised in the
Comment in a way that will help readers understand why one cannot discuss
quantum causes, not to mention other aspects of quantum theory, by simply
employing the tools of classical physics. Bell's inequalities have served a
very useful function in quantum foundations by showing how classical ideas,
which may seem plausible but disagree with von Neumann's Hilbert space
formulation of the theory, can lead to results in disagreement with
experiments. The proper way to honor the memory of one of the outstanding
figures of 20th century physics, whose influence on quantum foundations studies
can hardly be overestimated, is not by continuing to insist on ideas which
later developments have shown to be inadequate, but instead follow Bell's
example of careful, serious criticism of sloppy thinking and arm-waving
explanations, which he wanted to replace with clear concepts and consistent
mathematics in order to produce a structure worthy of being considered a key
part of theoretical physics. A good example of what he was looking for, but
obviously had not found, can be seen in his devastating critique in one of his
last papers, \emph{Against Measurement} \cite{Bll901}, of various textbook
discussions of measurement that he considered totally inadequate. We of course
cannot know what his reaction might have been to the current CH formulation of
the measurement process. But surely he would have begun with a careful reading
of the relevant publications before going on to perhaps identify serious flaws,
adopt some of the ideas, make significant improvements, or replace the whole CH
approach with something better. This attitude, applied to CH and other quantum
interpretations, is very much needed if the foundations community is to emerge
from its current disagreements and disarray, and arrive at a coherent
understanding of quantum theory, showing that we have actually made significant
progress since the golden years of 1925-26, the centenary of which is fast
approaching.

\xb
\section*{Acknowledgements}
\xa

The author expresses his appreciation to Carnegie-Mellon University and its
Physics Department for continuing support of his activities as an emeritus
faculty member.

\xb
\end{document}